# Anisotropic excitonic photocurrent in $\beta - Ga_2O_3$

Darpan Verma,[1] Md Mohsinur Rahman Adnan[2], Sushovan Dhara,[2] Chris Sturm[3], Siddharth Rajan,[2] and Roberto C. Myers[1,2*]

[1)] *Department of Materials Science Engineering, The Ohio State University, Columbus 43210 USA*

[2)] *Department of Electrical and Computer Engineering, The Ohio State University, Columbus 43210, USA*

[3)] *Felix-Bloch-Institut für Festkörperphysik, Universität Leipzig, Linnéstr. 5, 04103 Leipzig, German*

*\*myers.1079@osu.edu*

**Polarization dependent photocurrent spectra are measured on a (001) $\beta - Ga_2O_3$ Schottky photodetector, where the linear polarization of light is rotated within the $ab$ plane. Three spectral peaks at $4.92\ eV$, $5.15\ eV$, and $5.44\ eV$ are observed that vary in intensity with the optical polarization direction. The peak transition energies are consistent with excitons previously reported in $\beta - Ga_2O_3$ due to interband transitions modified by the valence band $p$-orbital anisotropy and the electron-hole Coulombic attraction. The measured polarization-dependence of the photocurrent matches our predictions based on electromagnetic simulations of anisotropic absorption using the complex dielectric function tensor extracted from previous ellipsometry studies. These results illustrate the dominance of excitonic absorption and photocurrent in $\beta - Ga_2O_3$ both below and above the band gap, demonstrate a combined theoretical/experimental understanding of anisotropic photocarrier generation, and validate previous atomistic band structure calculations in this low-symmetry ultra-wide band gap semiconductor.**

The ultra-wide bandgap energy semiconductor $\beta - Ga_2O_3$ is being studied for a number of applications [1–11] including so-called solar-blind photodetectors, which are generally defined as devices whose photocurrent shows negligible response to visible energy photons and a strong response to deep ultraviolet (UV) photons.[1–12] The necessary figures of merit for such devices are not yet well-defined, e.g. the exact cut-off and cut-on photon energies ($E_{ph}$). In particular, for terrestrial application, the solar spectrum does not cut-off at the limit of human vision ($E_{ph}$>3.26 eV), and emits significant photon flux out to 4.0 eV, while above the atmosphere the solar photon flux remains significant to 4.43 eV.[13,14] The starting point for evaluation of a material for such application is that its optical absorption edge ($E_{edge}$), which is not necessarily equal to the band gap ($E_g$), occur at least for $E_{ph} > 4.5$ eV.[15] From this perspective, $\beta - Ga_2O_3$ is an appealing candidate material since its $E_g = 5.05\ eV$.[16] As $\beta - Ga_2O_3$ is an indirect band gap semiconductor, the interband absorption is phonon-mediated and therefore one might expect a slow rise in absorption with $E_{ph}$, however both theory and experiments have shown a clear strong rise in photon absorption including strong peak structures both below and above $E_g$.[17–21] These absorption features are clearly distinct from indirect interband absorption and instead arise from strong electron-hole Coulombic attraction (excitons), which lowers the energy for the optical transition by this binding energy, around 230 meV on average for $\beta - Ga_2O_3$.[22–24] Another complexity arises from the low symmetry (monoclinic, C2/m) of this crystal,[25] which causes a rotational anisotropy for $E_{edge}$ due to a four-fold splitting of the valence band degeneracies.[16–19,26–30] Depending on the input light polarization direction ($\vec{E}$) with respect to the crystallographic axes, $E_{edge}$ varies by more than $0.36\ eV$, which has been observed in both transmittance and reflectance at room temperature, see Table I.[26–30] Although $E_{edge}$ for $\vec{E}//$a and $\vec{E}//$c almost coincide, for $\vec{E}//$b, $E_{edge}$ lies at a higher energy, making it more selective for deep UV detection.

Furthmüller and Bechstedt[16] carried out an extensive density functional theory (DFT) study to predict the optical properties of $\beta - Ga_2O_3$. Their quasiparticle (QP) band structure calculations (Hedin's GW) show that due to the low symmetry of the crystal structure, the valence band O 2p-like states are split leading to the first four QP interband transitions taking place from $p_z$ ($cz$ is 13.8° about b), $p_x$ ($x||a$), mixed $p_x/p_z$, and $p_y$ ($y||b$) valence bands and the Ga 4s conduction band with energies of 5.04 eV, 5.29 eV, 5.54 eV and 5.62 eV, respectively, plotted as thick dashed lines in Fig. 1. Although the minimum QP $E_g$ of $\beta - Ga_2O_3$ is $5.05\ eV$, the actual measured $E_{edge}$ can occur as low as 4.55 eV, due to the electron-hole Coulombic interaction (excitons), Table I. To account for this, DFT must also include solutions to the two-body Bethe Salpeter equation (BSE), which lead to a reduction in the interband transition energies (by the exciton binding energies) which themselves may vary depending on the different valence bands involved. For the four lowest interband transitions, described above, the authors found exciton absorption peaks at $4.65\ eV$, $4.90\ eV$, $5.0 - 5.2\ eV$, and $5.50\ eV$, shown in Fig. 1 as thin arrows.[16] We will refer to these transitions as $X_c$, $X_a$, $X_{ac}$, and $X_b$ to highlight their eXcitonic nature and anisotropic hole orbital component with respect to the monoclinic crystal frame. Subtracting the QP transition energies from the exciton peak energies, Furthmüller and Bechstedt estimated the exciton binding energies shown in Fig. 1, labeled $E_{Xc}$, $E_{Xa}$, $E_{Xac,}$ and $E_{Xb}$, whose values vary from $120 - 540\ meV$.

The underlying phenomena of a crystal field split valence band and a large exciton binding energy, discussed above, explain the early ambiguity in the literature on the proper $E_g$ of $\beta - Ga_2O_3$, since its value is not directly revealed by measurement of $E_{edge}$. As the sub-band gap exciton absorption can dominate the optical transitions, an accurate experimental/theoretical understanding of its role in photocurrent production is needed to understand and optimize solar blind PDs and may also improve the understanding of minority-carrier physics and breakdown phenomena. Polarization-resolved photocurrent spectroscopy can be used to shed light on the underlying excitonic transitions responsible for photocarrier generation and test the validity of electronic structure calculations as it is a sensitive quantitative probe of below and above $E_g$ absorption. Furthermore, in the presence of a DC electric field (F), exciton absorption leads to an eXciton Franz Keldysh effect (XFK effect), that red shifts the absorption edge with F and may be used to spectrally detect the onset of dielectric breakdown.[21]

Here we present polarization dependent photoresponsivity spectra from a (001) $\beta-Ga_2O_3$ Schottky photodetector as a function of linear polarization angle of the incident monochromatic light. The data reveal three distinct photocurrent peaks corresponding with the $X_a$, $X_{ac}$, and $X_b$ excitons, both below and above $E_g$. The highest binding energy exciton ($X_a$) photocurrent shows a strong rotational anisotropy, while the $X_{ac}$, and $X_b$ show a weaker polarization dependence. The data are modeled by utilizing the dielectric tensor measured by Sturm et. al.,[18] and then numerically solving the electro-magnetic wave equation to predict the anisotropic absorption spectra in $\beta-Ga_2O_3$.[31] Photoresponsivity spectra are then estimated by assuming all photon absorption within the depletion region leads to collected photocarriers. A detailed comparison of the acquired and predicted photocurrent spectra is presented and used to compare with the DFT literature to test the accuracy of the BSE based calculations.

A schematic of the experimental setup is shown in Fig. 2 (a), the same used in previous studies.[21,32] The monoclinic unit cell of $\beta-Ga_2O_3$ consists of lattice vectors $a$, $b$ and $c$, where α = γ = 90°, and β = 103.7°.[25] The sample is oriented with respect to the Cartesian lab frame such that, $x \parallel a$, $y \parallel b$, which forces c to be rotated by 13.7° from z about the b (in ac plane). The polarizer angle ($\theta$) determines the linear polarization direction ($\vec{E}$) of the monochromatic light, which is then focused on the $\beta-Ga_2O_3$ Schottky diode, using a 40 × Al-coated reflective microscope objective (NA = 0.5). The effective spot size of the illuminated light is measured using a knife edge measurement to be $8 \pm 0.3\ \mu$. Polarization-dependent photocurrent spectra are collected at the electrode edge with the $\beta-Ga_2O_3$ Schottky diode reverse biased, such that a DC electric field (F) is applied along z, using a Keithley 2604B source meter unit ($S$). The photocurrent is pre-amplified (Stanford Research System, SRS 560) and readout using a digital lock-in amplifier (Zurich instruments HF2LI) referenced to the optical chopper modulating at 550 $Hz$ frequency. Relative to other photocurrent studies in $\beta-Ga_2O_3$, we select a high modulation frequency to avoid the photo-carrier gain effect that occurs at $> 100\ ms$ timescales.[33,34] The output power ($W$) is measured at each $E_{ph}$ and $\theta$ using a wavelength corrected power meter (Thorlabs PM100D) placed in front of the microscope objective to convert the raw measured photocurrent (A) into polarization-dependent responsivity ($A/W$).

The band edge diagram for the Schottky diodes used in this study is shown in Fig 2(b), which are fabricated on a hydride vapor phase epitaxy (HVPE) $10\ \mu m$ thick $\beta - Ga_2O_3$ epitaxial layer grown on a 0.65 mm thick, n+ (001) $\beta - Ga_2O_3$ substrate (Tamura Corporation). The $100\ \mu m$ circular top Schottky contacts are patterned using optical lithography followed by electron beam evaporation of $Pt/Au$ ($30/100\ nm$), and a $Ti/Au$ ($30/100\ nm$) layer is blanket deposited to serve as an ohmic contact. The net doping concentration ($N_d - N_a$) of the epitaxial layer is $1.37 \times 10^{16}\ cm^{-3}$. The band diagram is obtained using a one-dimensional self-consistent Poisson solver.[35,36] The built-in depletion barrier of $400\ nm$ is responsible for photocarrier collection at $0\ V$ bias. The $I - V$ characteristics taken in the dark, yield a reverse leakage current density of $<\ 1\ nA/cm^2$ at 10 V, and the device showed destructive breakdown at 285 V.

Figures 2(c) and (d) plot the measured polarization-dependent photocurrent spectra on a polar intensity graph for 0, and $40\ V$ reverse bias, respectively. Color intensity corresponds to photoresponsivity amplitude ($I_{PR}$), while the radial and angular coordinates correspond to $E_{ph}$ and $\theta$, respectively. $E_{edge}$ of the photocurrent varies for $\vec{E}||a$ and $\vec{E}||b$ from $4.63 \pm 0.004\ eV$ to $4.89 \pm 0.006\ eV$ for $0\ V$ reverse bias data respectively. Similarly, for $40\ V$ reverse bias data it varies from $4.60 \pm 0.005\ eV$ to $4.73 \pm 0.006\ eV$, respectively. In comparison to previous experimental studies (Table I), our 0 V $E_{edge}$ for $\vec{E}||b$ is close to that of Matsumoto et. al.[26] signifying that the selection rule in our case permits the transition from $\Gamma_1^-(2)$ to $\Gamma_1^+(1)$, instead of $\Gamma_1^-(1)$ to $\Gamma_1^+(1)$.[28]

Polarization dependent photocurrent spectra are modeled by solving the electro-magnetic (EM) wave propagation decay within a thick slab of $\beta - Ga_2O_3$ using the measured complex dielectric function tensor ($\boldsymbol{\varepsilon}$) by Sturm et. al.[18] As the EM wave propagates into the $\beta - Ga_2O_3$ slab it decays due to absorption.[31] A detailed discussion of the numerical modeling of the EM wave propagation is in preparation.[31] Assuming that all photon absorption results in collected photocarriers, the photocurrent spectra are numerically modeled by integrating the generation rate across the photocarrier collection region, assumed to be equal to the depletion region. At 0 V and 40 V, the depletion widths correspond to 400 nm and 1800 nm, respectively. Figures 2 (e) and (f) plot the numerically modeled spectra as a function of polarization angle.

Line cuts of the polar intensity graphs in Figs. 2 (c-f) for $\theta = 0, 90$ ($\vec{E} \parallel a, b$), i.e., light polarized along either of the two principal axes, are plotted in Figs. 2(g) and (h). The measured responsivity spectra are well fit by three peaks, Fig. 2(i), which we ascribe to the three exciton peaks ($X_a$, $X_{ac}$ and $X_b$), discussed previously and shown in Fig. 1. The measured (symbols) and simulated (line) photocurrent spectra at 0 V and 400 nm collection region (Fig 2(g)) show that the $X_a$ ($\vec{E} \parallel a$) and $X_b$ ($\vec{E} \parallel b$) peaks are captured in the modeled photocurrent spectra. Although the $X_{ac}$ peak is clearly observed in the data, it is not predicted in the model, which is expected since the transition was predicted to be quite weak.[16] Its presence in the photocurrent spectra could be related to the high NA of the microscope objective leading to some out-of-plane components to $\vec{E}$, which might excite the mixed $X_{ac}$ transitions, or due to a possibly enhanced dissociate rate compared to the other excitons. At 40 V (Fig 2(h)), the modeled spectra show the $X_a$ and $X_{ac}$ peaks, but not $X_b$. As the model contains no XFK effect, the bias dependence in the model is based entirely on the variation of the depletion width, which changes the photocurrent contribution of different transition energies. Wider depletion widths at higher bias increase the contribution of weaker transitions (with longer absorption depths), which explains the disappearance of the high energy $X_b$ and appearance of the weaker lower energy $X_{ac}$ peak.

We carry out three peak fits of the spectra to determine their minimum center peak position ($E_{ph}^0$) at any value of $\theta$, which are 4.92 $eV$ ($X_a$), 5.15 $eV$ ($X_{ac}$) and 5.44 $eV$ ($X_b$) at 0 V. These values are in close agreement with the BSE based calculations of Furthmüller and Bechstedt, with the notable absence of the $X_C$ exciton at 4.65 eV in our data. This absence is expected due to the (001) orientation of the substrate, which under normal incidence greatly reduces (but does not fully eliminate) the c component of $\vec{E}$.

Noting that the exciton binding energy, by definition, cannot vary with $\theta$ as it is a scalar, three peak fits are carried out on the polarization dependent photoresponsivity spectra holding the three values of $E_{ph}^0$ constant and allowing only the amplitudes and widths of the peaks to vary to obtain the best spectral fit at every value of $\theta$. The results are normalized and plotted in Fig. 3(a) and 3(b) showing that the rotational anisotropy of the photocurrent is dominated by the $X_a$ transition showing its largest magnitude for $\theta = 0$ ($\vec{E} \parallel a$), and strongly suppressed for $\theta = 90$ ($\vec{E} \parallel b$) for 0V and 40 V responsivity spectra. In comparison, the $X_{ac}$ and $X_b$ amplitudes are almost isotropic. This is perhaps related to carrier screening effects that may take place for the $X_{ac}$ and $X_b$ that occur at $E_{ph} > E_g$, versus $E_{ph} < E_g$ transitions $X_a$. Line cuts taken from the modeled normalized photoresponsivity at the values of experimentally observed central energy $E_{ph}^0$ (4.92 eV, 5.15 eV and 5.44 eV) for 400 nm and at $E_{ph}^0$ (4.85 eV, 5.14 eV and 5.47 eV) for 1800 nm are plotted in Fig 3 (c) and (d) for $X_a$, $X_{ac}$, and $X_b$, respectively. The modeled photocurrent polarization anisotropy shows a similar a-polar anisotropy for the $X_a$ peak as observed in the data and the $X_b$ transition showed no anisotropy at any depletion width in the modeled data, in agreement with the data. However, for the near band gap $X_{ac}$ peak, the model shows a cross-over behavior where the narrow depletion width model predicts an a-polar $X_{ac}$ and an isotropic behavior for wider depletion. As previously discussed above, the $X_{ac}$ peak observed in photocurrent displays an anomalously large contribution to the measured photocurrent compared to what is predicted suggesting that a more accurate model would need to take into account the variation in dissociation rates for the different excitons to better capture the photocurrent polarization dependence. As previously discussed above, Furthmüller and Bechstedt numerically estimated the binding energies ($E_X$) by comparing the BSE results to the QP calculations, Fig. 1. Using the same comparison to the QP results, the photocurrent peak positions in our study yield $E_{X_a}$ = 0.37 eV, $E_{X_{ac}}$ = 0.39 eV, and $E_{X_b}$ = 0.18 eV. These results agree well with Ref. [16].

Bias dependent measurements of the photoresponsivity spectra for $\vec{E}||b$ and $\vec{E}||a$ were carried out at a different position on the same sample, Fig 4 (a) and (b), respectively. The insets of Fig 4(a) and (b) show the three peaks fits for $X_a$, $X_{ac}$ and $X_b$ color coded in the same way as in Figs. 1, 2 and 3. The bias dependent data show negligible peak shifting, indicating that the electric field is well below the breakdown field for the measured biases, limited to -40 V. This agrees with previous measurements where an XFK red shift begins to be noticed at electric fields above 1 MV/cm.[21] Therefore, the bias dependent data are fitted while keeping the peak energies constant and varying only the peak amplitudes ($I_{ph}^0$). The raw (not normalized) photocurrent peak magnitudes are plotted as a function of bias for each of the three excitonic peaks in Figs. 4 (c-e) in order to allow quantitative comparison of their magnitudes. For each peak, data were taken with the light polarized $\vec{E}||a$ (left vertical axes) and $\vec{E}||b$ (right vertical axes), which are plotted on separate scales in order to compare their bias dependences. Examining Fig. 4 (c), the $X_a$ exciton photocurrent is strongly anisotropic, about 6 times large for $\vec{E}||a$ than for $\vec{E}||b$, as previously shown in Fig. 3. Although the overall magnitudes are different, the relative bias dependence of the photocurrent is identical in both polarization configurations. This is expected since the linear polarization angle only affects the magnitude of the absorption, but does not impact the bias dependent exciton ionization rate. The results are also consistent with the low field region in which the quantum efficiency is limited by free exciton dissociation,[21] such that the photocurrent increases with bias. Similar results are obtained for $X_{ac}$, where in Fig. 4 (d) the bias dependences show the same increasing trend, albeit with larger data scatter for $\vec{E}||a$ particularly at lower biases. On the other hand, Fig. 4 (e) shows that the bias dependence of $X_b$ exhibits a clearly distinct variation between $\vec{E}||a$ and $\vec{E}||b$ polarized excitation; one rises while the other decreases with bias. The results can be interpreted by noting that the $X_b$ transition, being well above band gap, should experience photocarrier density dependent screening effects. That is, $X_b$ excitons are excited with photons with energies well above the bandgap, such that $X_b$ are generated along with a large population of photocarriers. Thus, the e-h Coulombic potential that forms $X_b$ would be screened,[24] and enable more efficient exciton dissociation. In this case, one would expect a negligible bias dependence to the dissociation rate and resulting photocurrent as $X_b$ are easily dissociated. But simply invoking screening does not explain the apparent anisotropy to the bias dependence for the case of $X_b$. Additional power dependent photocurrent measurements may be needed to explore density dependent screening effects and shed light on its polarization dependence.

The simulation results, Fig. 4 (f) and (g), show that with increasing slab collection region (increasing depletion width with bias), not only is there a relative rise in amplitude but also there is an apparent redshift to the photocurrent spectra regardless of polarization angle. As described above, the model does not include any XFK physics, and therefore the redshift seen here is simply due to the changing photocurrent contribution as the depletion width varies, which at high bias favors the weaker (lower energy) transitions. As previously discussed, the simple photocurrent model assumes all carriers within the depletion width are collected, which ignores details of photocarrier collection. The discrepancy between the bias dependent model and data therefore reveal that there is a spectral dependence to the photocarrier collection efficiency. Such a photon energy dependence would not be expected in the absence of excitonic effects, barring unusual intervalley scattering processes.

To summarize, three distinct photocurrent peaks are observed in the polarization-resolved photocurrent spectra of $\beta - Ga_2O_3$ (001) at room temperature, whose magnitudes vary with linear polarization direction with respect to the crystal. These peaks quantitatively match the absorption peak energies of Furthmüller and Bechstedt, thus validating their estimates based on BSE and QP band structure calculations.[16] For (001) oriented crystal, we observe photocurrent peaks corresponding to $X_a$, $X_{ac}$, and $X_b$ excitonic transitions, but not $X_c$, which are labeled to indicate the O 2p-like orbital character of the crystal field split valence bands from which they originate. Strong rotational-anisotropy of the photocurrent magnitude is observed for the strongest binding energy exciton ($X_a$), while the $X_{ac}$, and $X_b$ are much more isotropic. A polarization-dependent photocurrent spectral model is presented based on the complex dielectric function of $\beta - Ga_2O_3$,[18] which calculates EM solutions for polarized light propagating in this monoclinic crystal.[31] As a first approximation, we assume that all photon absorption leads to collected photocarriers within the depletion region (ignoring more complex minority carrier limitations and higher order optical processes). This model is able to capture the polar anisotropy and peak structure of the measured photocurrent spectra. However, the spectral and field dependences are not entirely matched to the model. Namely, the $X_{ac}$ transition is expected to be too weak to be observed, yet appears prominently for the (001) orientation. Additionally, the above band gap photocurrent peak ($X_b$) shows an unusual bias dependence which varies with the optical polarization axis. The differences in the spectral lineshape between the modeled and measured photocurrent imply an energy dependence to the photocarrier collection process, which is not explained by simply the excitonic absorption process, which would have been captured by the model. This strong spectral variation in the above band gap optoelectronic response of $\beta - Ga_2O_3$ (001), and its non-trivial bias dependence provide striking evidence for the breakdown of free photocarrier drift-diffusion models. Further studies are needed to model and measure anisotropic optoelectronic processes in low symmetry semiconductors, particularly the UWBG systems in which electron-hole interactions dominate carrier generation and collection processes. Ultimately, these phenomena may provide new optical and electronic sensing functionalities based on anisotropic excitonic photocurrents.

**Acknowledgments** Funding for this research was provided by the Center for Emergent Materials: an NSF MRSEC under award number DMR-1420451 and by the AFOSR GAME MURI (Grant FA9550-18-1-0479, Program Manager Dr. Ali Sayir).

## References

[1] M. Chang, J. Ye, Y. Su, J. Shen, N. Zhao, J. Wang, H. Song, X. Zhong, S. Wang, W. Tang, and D. Guo, J Phys D Appl Phys **55**, 035103 (2022).

[2] Y. Qin, L.H. Li, Z. Yu, F. Wu, D. Dong, W. Guo, Z. Zhang, J.H. Yuan, K.H. Xue, X. Miao, and S. Long, Adv Sci **8**, 2101106 (2021).

[3] Y. Wang, H. Li, J. Cao, J. Shen, Q. Zhang, Y. Yang, Z. Dong, T. Zhou, Y. Zhang, W. Tang, and Z. Wu, ACS Nano **15**, 16654 (2021).

[4] Y. Wang, Z. Yang, H. Li, S. Li, Y. Zhi, Z. Yan, X. Huang, X. Wei, W. Tang, and Z. Wu, ACS Appl Mater Interfaces **12**, 47714 (2020).

[5] S. Kim, S. Oh, and J. Kim, ACS Photonics **6**, 1026 (2019).

[6] Y. Qin, S. Long, Q. He, H. Dong, G. Jian, Y. Zhang, X. Hou, P. Tan, Z. Zhang, Y. Lu, C. Shan, J. Wang, W. Hu, H. Lv, Q. Liu, and M. Liu, Adv Electron Mater **5**, 1900389 (2019).

[7] P. Jaiswal, U. Ul Muazzam, A.S. Pratiyush, N. Mohan, S. Raghavan, R. Muralidharan, S.A. Shivashankar, and D.N. Nath, Appl Phys Lett **112**, 021105 (2018).

[8] S. Oh, C.K. Kim, and J. Kim, ACS Photonics **5**, 1123 (2018).

[9] W.E. Mahmoud, Sol Energy Mater Sol Cells **152**, 65 (2016).

[10] W.Y. Kong, G.A. Wu, K.Y. Wang, T.F. Zhang, Y.F. Zou, D.D. Wang, and L.B. Luo, Adv Mater **28**, 10725 (2016).

[11] Z. Ji, J. Du, J. Fan, and W. Wang, Opt Mater **28**, 415 (2006).

[12] G. Zeng, X.X. Li, Y.C. Li, D.B. Chen, Y.C. Chen, X.F. Zhao, N. Chen, T.Y. Wang, D.W. Zhang, and H.L. Lu, ACS Appl Mater Interfaces **14**, 16846 (2022).

[13] L.X. Qian, W. Li, Z. Gu, J. Tian, X. Huang, P.T. Lai, and W. Zhang, Adv Opt Mater **10**, 2102055 (2022).

[14] X. Xia, J.-S. Li, R. Sharma, F. Ren, M.A.J. Rasel, S. Stepanoff, N. Al-Mamun, A. Haque, D.E. Wolfe, S. Modak, L. Chernyak, M.E. Law, A. Khachatrian, and S.J. Pearton, ECS J Solid State Sci Technol **11**, 095001 (2022).

[15] X. Chen, F. Ren, S. Gu, and J. Ye, Photonics Res **7**, 381 (2019).

[16] J. Furthmüller and F. Bechstedt, Phys Rev B **93**, 115204 (2016).

[17] J.B. Varley and A. Schleife, Semicond Sci Technol **30**, 024010 (2015).

[18] C. Sturm, J. Furthmüller, F. Bechstedt, R. Schmidt-Grund, and M. Grundmann, APL Mater **3**, 106106 (2015).

[19] C. Sturm, R. Schmidt-Grund, C. Kranert, J. Furthmüller, F. Bechstedt, and M. Grundmann, Phys Rev B **94**, 035148 (2016).

[20] A. Mock, R. Korlacki, C. Briley, V. Darakchieva, B. Monemar, Y. Kumagai, K. Goto, M. Higashiwaki, and M. Schubert, Phys Rev B **96**, 245205 (2017).

[21] M.M.R. Adnan, D. Verma, Z. Xia, N.K. Kalarickal, S. Rajan, and R.C. Myers, Phys Rev Appl **16**, 034011 (2021).

[22] T. Onuma, S. Saito, K. Sasaki, K. Goto, T. Masui, T. Yamaguchi, T. Honda, A. Kuramata, and M. Higashiwaki, Appl Phys Lett **108**, 101904 (2016).

[23] T. Onuma, K. Tanaka, K. Sasaki, T. Yamaguchi, T. Honda, A. Kuramata, S. Yamakoshi, and M. Higashiwaki, Appl Phys Lett **115**, 231102 (2019).

[24] F. Bechstedt and J. Furthmüller, Appl Phys Lett **114**, 122101 (2019).

[25] S. Geller, J Chem Phys **33**, 676 (1960).

[26] T. Matsumoto, M. Aoki, A. Kinoshita, and T. Aono, Jpn J Appl Phys **13**, 1578 (1974).

[27] F. Ricci, F. Boschi, A. Baraldi, A. Filippetti, M. Higashiwaki, A. Kuramata, V. Fiorentini, and R. Fornari, J Phys: Condens Matter **28**, 224005 (2016).

[28] T. Onuma, S. Saito, K. Sasaki, T. Masui, T. Yamaguchi, T. Honda, and M. Higashiwaki, Jpn J Appl Phys **54**, 112601 (2015).

[29] N. Ueda, H. Hosono, R. Waseda, and H. Kawazoe, Appl Phys Lett **71**, 933 (1997).

[30] X. Chen, W. Mu, Y. Xu, B. Fu, Z. Jia, F.F. Ren, S. Gu, R. Zhang, Y. Zheng, X. Tao, and J. Ye, ACS Appl Mater Interfaces **11**, 7131 (2019).

[31] M.M.R. Adnan, D. Verma, C. Sturm, and R. Myers, *(in preparation)*.

[32] D. Verma, M.M.R. Adnan, M.W. Rahman, S. Rajan, and R.C. Myers, Appl Phys Lett **116**, 202102 (2020).

[33] A.S. Pratiyush, S. Krishnamoorthy, R. Muralidharan, S. Rajan, and D.N. Nath, in *Advances in $Ga_2O_3$ Solar-Blind UV Photodetectors, in: Gallium Oxide* (Elsevier, 2018), pp. 369–399.

[34] A. Singh Pratiyush, S. Krishnamoorthy, S. Vishnu Solanke, Z. Xia, R. Muralidharan, S. Rajan, and D.N. Nath, Appl Phys Lett **110**, 221107 (2017).

[35] M. Grundmann, BandEng: Poisson-Schrodinger Solver Software, https://my.ece.ucsb.edu/mgrundmann/bandeng/ (2004).

[36] B. Jogai, J Appl Phys **91**, 3721 (2002).

# Tables

TABLE I. Measured $E_{edge}$ variation with linear polarization direction ($\vec{E}$) from transmission (reflectance) measurements (in eV). Note that a* is $-13.7^0$ from a and perpendicular to (100).

| References | $\vec{E}\|\|a$ | $\vec{E}\|\|a*$ | $\vec{E}\|\|[102]$ | $\vec{E}\|\|b$ | $\vec{E}\|\|c$ |
|---|---|---|---|---|---|
| [26] | | 4.56 | 4.56 | 4.90 (5.06) | 4.54 (4.63) |
| [29] | | | | 4.79 | 4.52 |
| [28] Mg- doped | | 4.57 (4.58) | | | 4.48 |
| [28] undoped | | 4.55 (4.54) | | 4.70 (4.73, 4.73, 4.88) | |
| [30] | | | | 4.76 | 4.53 |
| [27] | | 4.57 | 4.52 | 4.72 | 4.54 |
| This work (0 V) | 4.63 | | | 4.89 | |
| This work (40V) | 4.60 | | | 4.73 | |

**Figures & Figure Captions**

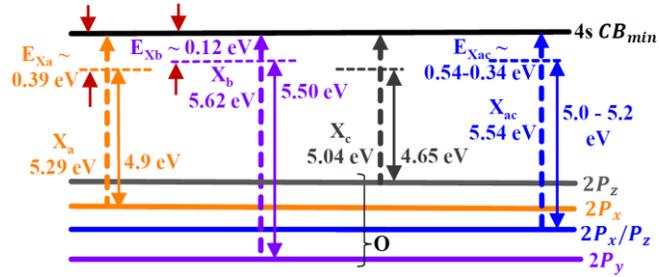

Fig. 1. Interband optical transitions in $\beta - Ga_2O_3$ (thick dashed lines) and the exciton reduced transitions (thin lines) as calculated by Furthmüller and Bechstedt.[16]

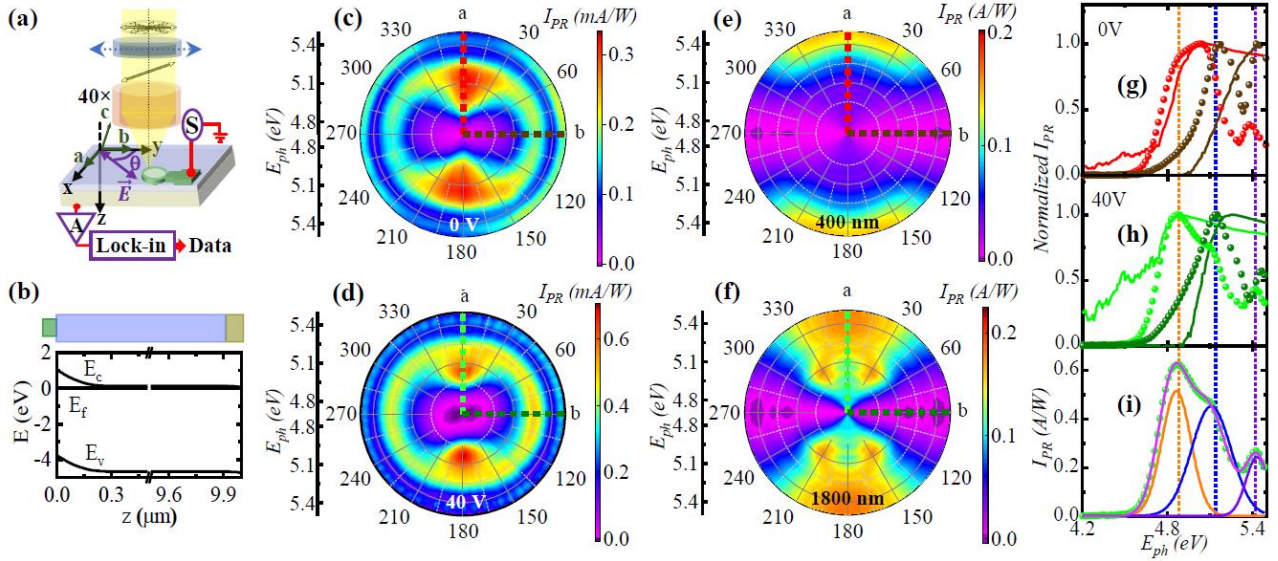

Fig. 2. (a) Geometry of polarization-resolved photocurrent spectroscopy setup, with optical E-field vector ($\vec{E}$) at angle $\theta$ and $k_{ph}//z$. The (001) oriented $\beta - Ga_2O_3$ crystal is aligned such that $a||x$, $b||y$, causing c to lie 13.2° from $-z$ about b (within the ac plane). (b) Band edge diagram of the diode. (c)-(f) Spectral and polarization dependent photoresponsivity, $I_{PR}$ color intensity. The radial axis is $E_{ph}$ and $\theta$ the azimuthal. (c), (d) Experimentally measured polarization dependent $I_{PR}$ spectra at two different biases, $0\ V$ and $40\ V$, respectively, and (e),(f) simulated spectra calculated for depletion widths of $400\ nm$ and $1800\ nm$ corresponding to biases of 0 and $40\ V$, respectively. (g), (h) Line cuts (color coded) from the 3D data sets (polar plots c-f) for polarization along the principle axes. Measured data are plotted as points and calculations as lines. (i) Representative three peak fit, $X_a$ (orange), $X_{ac}$ (blue), and $X_b$ (violet) for 40 V data with $\vec{E}||a$. The cumulative fitted spectrum is plotted in magenta.

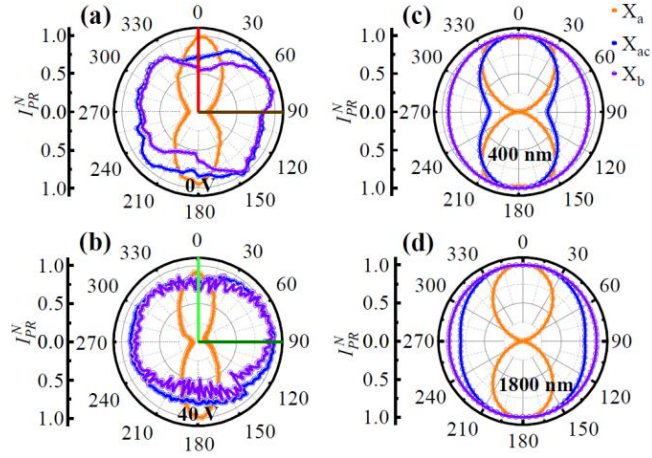

Fig. 3. Rotational anisotropy of the excitonic photocurrent in $\beta - Ga_2O_3$. (a),(b) $\theta$ dependence of the photocurrent peaks normalized by their maximum values, $I_{PR}^N = I_{PR}^0/[\max of\ I_{PR}^0(\theta)]$, for data measured at 0V and 40 V, respectively, based on three peak fits of the data in Fig. 2. (c) and (d) Polarization dependence of the simulated photocurrent spectra for depletion widths of $400\ nm$ and $1800\ nm$ corresponding to biases of 0 and 40 $V$, respectively, from the photocurrent amplitude at the peak position for the three excitonic transitions, and normalized by their maximum value.

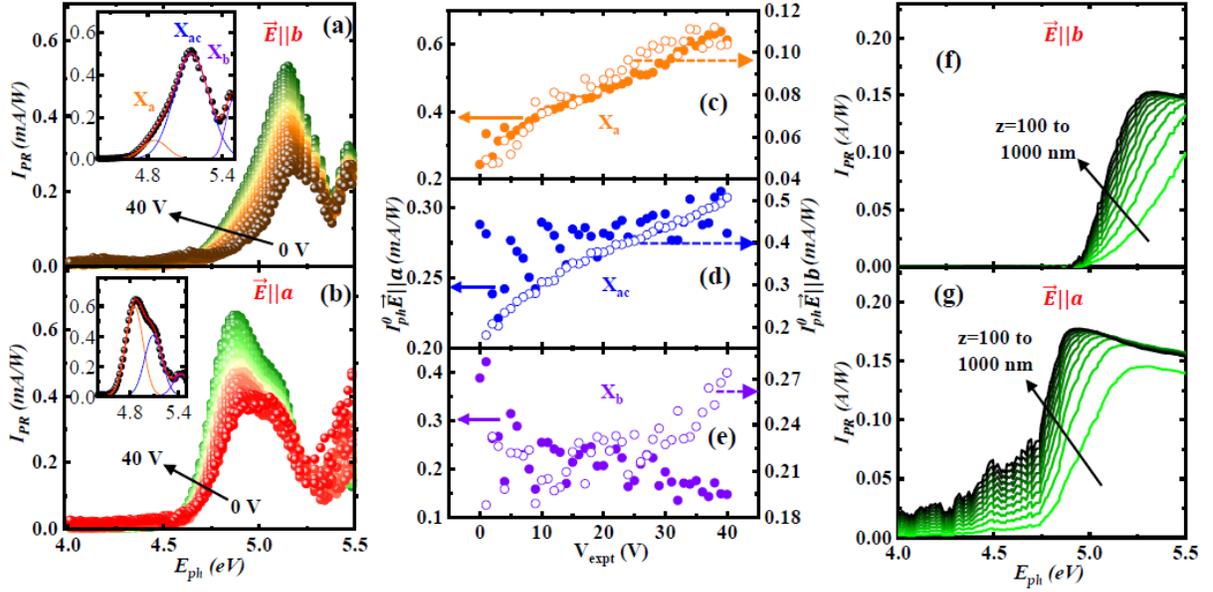

Fig. 4. (a),(b) Bias dependent photocurrent spectra for $\vec{E}||b$ and $\vec{E}||a$, respectively. Insets show representative three peak fits at 40V reverse bias. Exciton peaks are color coded as in Fig. 1. (c)-(e) Bias dependence of the photocurrent peak amplitudes ($I_{PR}^0$) from three peak fits of the data in (a) and (b). The left and right vertical axes correspond to light polarized along $\vec{E}||a$ and $\vec{E}||b$, respectively. (f),(g) Simulated $I_{PR}$ spectra are plotted at various depletion widths for $\vec{E}||b$ and $\vec{E}||a$, respectively.

15